\documentclass[structabstract]{aa}  
\usepackage{graphicx}
\usepackage{natbib}
\bibpunct{(}{)}{;}{a}{}{,} 
\usepackage{amsmath}

\begin{document}
   \title{Confirmation of 1RXS J165443.5$-$191620 as an intermediate polar and its orbital and spin periods}
\titlerunning{The IP RXJ1654}
\author{S. Scaringi\thanks{e-mail: s.scaringi@astro.ru.nl}\inst{1}
          \and
          S. Connolly\inst{2}
	  \and
	  J. Patterson\inst{3}
	  \and
	  J. R. Thorstensen\inst{4}
	  \and
	  H. Uthas\inst{2,3}
	  \and
	  C. Knigge\inst{2}
	  \and
	  L. Vican\inst{3}
	  \and
	  B. Monard\inst{5}
	  \and
	  R. Rea\inst{6}
	  \and
	  T. Krajci\inst{7}
	  \and
	  S. Lowther\inst{8}
	  \and
	  G. Myers\inst{9}
	  \and
	  G. Bolt\inst{10}
	  \and
	  A. Dieball\inst{2}
	  \and
	  P.J. Groot\inst{1}
          }

   \institute{Department of Astrophysics, Radboud University Nijmegen, P.O. Box 9010, 6500 GL Nijmegen, The Netherlands\\
         \and
	  Department of Physics and Astronomy, University of Southampton, Highfield, Southampton, SO17 1BJ, UK\\
         \and
	  Department of Astronomy, Columbia University, 550 West 120th Street, New York, NY 10027, USA\\
         \and
	  Department of Physics and Astronomy, Dartmouth College, 6127 Wilder Laboratory, Hanover, NH 03755-3528, USA\\
         \and
	  CBA Pretoria, PO Box 11426, Tiegerpoort 0056, South Africa\\
         \and
	  CBA Nelson, 8 Regent Lane, Richmond, Nelson, New Zealand\\
         \and
	  CBA New Mexico, PO Box 1351, Cloudcroft, NM, 88317, USA\\
         \and
	  CBA Pukekohe, 19 Cape Vista Crescent, Pukekohe 2120, New Zealand\\
         \and
	  CBA San Mateo, 5 Inverness Way, Hillsborough, CA, 94010, USA\\
         \and
	  CBA Perth, 295 Camberwarra Drive, Craigie, Western Australia 6025\\
             }



   \date{Received September ...; accepted March ...}

 
  \abstract
   {}
   {We investigate the physical nature of the X-ray emitting source 1RXS J165443.5$-$191620 through optical photometry and time-resolved spectroscopy.}
   {Optical photometry is obtained from a variety of telescopes all over the world spanning $\approx$27 days. Additionally, time-resolved spectroscopy is obtained from the MDM observatory.}
   {The optical photometry clearly displays modulations consistent with those observed in magnetic cataclysmic variables: a low-frequency signal interpreted as the orbital period, a high-frequency signal interpreted as the white dwarf spin period, and an orbital sideband modulation. Our findings and interpretations are further confirmed through optical, time-resolved, spectroscopy that displays H$\alpha$ radial velocity shifts modulated on the binary orbital period.}
   {We confirm the true nature of 1RXS J165443.5$-$191620 as an intermediate polar with a spin period of 546 seconds and an orbital period of 3.7 hours. In particular, 1RXS J165443.5$-$191620 is part of a growing subset of intermediate polars, all displaying hard X-ray emission above 15keV, white dwarf spin periods below 30 minutes, and spin-to-orbital ratios below 0.1.}


   \maketitle
%

\section{Introduction}

Cataclysmic variables (CVs) are close binary systems consisting of a late-type star transferring material onto a white dwarf (WD) companion via Roche-lobe overflow, and typically have orbital periods of the order of hours. Mass transfer is driven by the loss of orbital angular momentum, which in turn is thought to be caused by magnetic braking of the secondary star, except at the shortest periods ($\sim$ 2 hours or less), where gravitational radiation dominates. This mass transfer leads to the formation of an accretion disc subject to various instabilities, which can then produce outburst events such as dwarf novae (\citealt{warner})

Magnetic CVs (mCVs) are a subset of the catalogued CVs ($\approx 10\% - 20\%$, \citealt{downes}; \citealt{RKcat}), and generally fall into two categories: polars (or AM Her types after the prototype system) and intermediate polars (IPs or DQ Her types). The WDs in polars have sufficiently strong magnetic fields ($10^7 - 10^9$ Gauss) to prevent the formation of an accretion disc and to lock the secondary late-type star, synchronising the whole system (for a review of polars, see \citealt{cropper}). The strong magnetic field in these systems is confirmed by strong linear and circular polarisation, together with measurements of cyclotron humps (\citealt{warner}). For IPs, the lack of strong optical polarisation implies a weaker magnetic field, which is not powerful enough to synchronise the secondary star (for a review of IPs, see \citealt{patterson}). In these systems, material leaving the $L_{1}$ point only forms an accretion disc up to the point where the magnetic pressure exceeds the ram pressure of the accreting gas. From this point onwards the accretion dynamics are governed by the magnetic field lines, which channel the material onto the WD magnetic poles. The nature of these systems is confirmed by the detection of coherent X-ray and/or optical modulations associated with the spin period of the WD. It is however not uncommon for the dominant optical frequency to differ from the X-ray frequency.  When this occurs the X-ray signal is usually associated to white dwarf spin frequency ($\omega$), and the optical frequency is most often observed to be $\omega - \Omega$, where $\Omega$ is the orbital frequency. These optical modulations are thought to arise from X-ray reprocessing within the surrounding medium, where the X-ray emission from the WD pole illuminates an unknown structure fixed in the reference frame rotating with the binary (\citealt{warner81}). Other ``orbital side bands'' (e.g. $\omega + \Omega$) can also be produced, and have been detected in some IPs as well (e.g. \citealt{warner86}).

Hard X-ray surveys such as the {\it INTEGRAL}/IBIS survey (\citealt{cat4}) have proven remarkably efficient in detecting mCVs, and in particular intermediate polars with spin-to-orbital ratios below 0.1. (\citealt{scaringi,barlow}). The low and persistent hard X-ray flux of IPs above 15 keV was also expected because previous X-ray spectroscopy at lower energies revealed a hard X-ray excess (\citealt{lamb,chanmugam}).

Here we report on the optical photometry and time-resolved spectroscopy of one candidate IP, 1RXS J165443.5$-$191620  (hereafter RXJ1654). This \textit{Rosat} detection has been also catalogued in the {\it INTEGRAL}/IBIS survey (\citealt{cat4}); but remained unidentified until a tentative classification by \cite{masettiVIII} based on optical (follow-up) spectroscopy. The spectra displayed clear Balmer and HeI emission lines, with HeII~$\lambda$4686\AA/H$\beta$ equivalent width ratios $\ge0.5$, suggesting an intermediate polar classification for this CV. Recently \cite{lutovinov} have obtained optical photometry for RXJ1654 that displays clear modulation of $\approx 550$ seconds, consistent with that of the WD spin, which enforces the tentative classification of this object as an IP, but does not confirming its true nature. \cite{pretorius} has already discussed the dangers of inferring IP classifications based on the optical spectra and detection above 15 keV alone. Out of a sample of five hard X-ray emitting systems with tentative IP classification from optical spectroscopy, three showed clear signs of orbital and spin modulations, whilst two did not.

Strong corroboration of the suspected IP status can come from long time-span optical photometry showing the expected sideband structure, and time-resolved radial velocity measurements showing periodic variations on the orbital period. Photometry yields incontrovertible orbital periods when eclipses are present, but other modulations can masquerade as orbital periods as well. Time-resolved spectroscopy is thus the most reliable technique for determining orbital periods of CVs. This together with the detection of the $\omega - \Omega$ frequency, and/or coherent X-ray modulations consistent with $\omega$ (\citealt{buckley00}), would then unambiguously confirm the nature of an IP, excluding any other origin to the observed photometric variations. 

In Section~2 we will present preliminary observations obtained with the IAC80 during the university of Southampton student field trip. The data displayed a candidate WD spin modulation, which then motivated us towards obtaining a better sampled lightcurve spanning a wider range in time for this system to search for orbital periodicities. Section~3 describes the effort and data analysis to achieve this on a variety of telescopes with the global network of astronomers part of the Center for Backyard Astrophysics (CBA). In this respect, we confirm and update the \cite{lutovinov} candidate spin period, and also discuss the detection of the orbital period and the $\omega - \Omega$ signal. Additional time-resolved spectroscopy was also obtained on the Hiltner Telescope at MDM Observatory and this is described in Section~4, together with the data analysis and clear H$\alpha$ radial velocity shifts associated to the orbital period, which undoubtedly confirms the class membership of RXJ1654 as an IP. 


\section{IAC80 preliminary observations}
\begin{figure}
\includegraphics[width=0.5\textwidth]{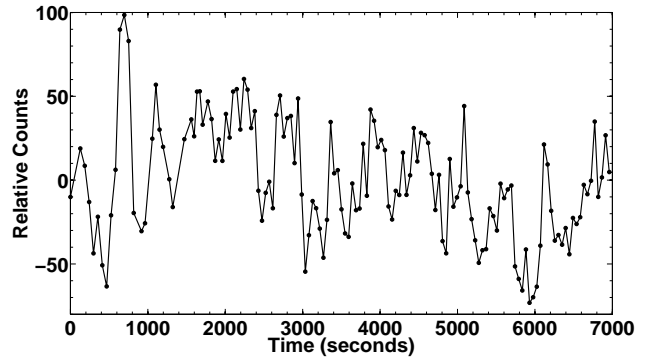}	
\caption{IAC80 differential photometry lightcurve with $\approx$60 second cadence on the source RXJ1654}
\label{fig:1}
\end{figure}

\begin{figure}
\includegraphics[width=0.5\textwidth]{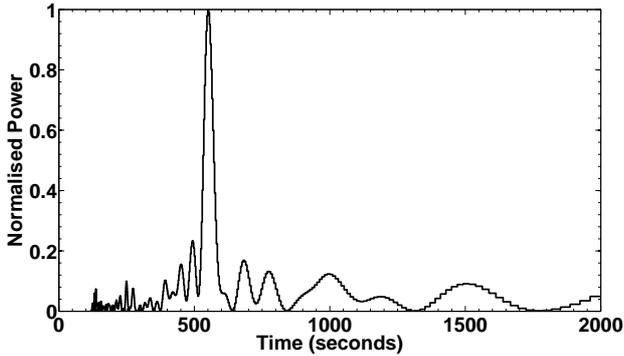}	
\caption{RXJ1654 power spectrum, showing a clear signal at $\approx$550 seconds.}
\label{fig:2}
\end{figure}

$V$-band photometry on RXJ1654 was obtained with the IAC80 80 cm cassegrain reflector telescope during the night of March 31 2010, with a 2048x2048 pixel CCD yielding a plate scale of 0.33 arcseconds/pixel. This was part of an observing project carried out during an undergraduate astronomy field trip to the Observatorio del Teide organised by the university of Southampton. The motivation for the project was the tentative spectroscopic classification of RXJ1654 as an mCV by \cite{masettiVIII}. The source was observed for a total of about 2 hours, with variable exposures between 20 and 40 seconds to compensate for variable atmospheric conditions during the observation. A read-out time of 22 seconds allowed us to achieve a lightcurve for the source with $\approx$ 60 second cadence. The observations were de-biased and flat-fielded with calibration frames obtained at the beginning and at the end of the night. Differential aperture photometry was then performed on RXJ1654 using three reference stars within the field. The differential magnitude lightcurve from the observation can be found in Fig. \ref{fig:1}. Even though the length of the observation was not very long, sinusoidal variations of the order of a few hundred seconds are visible in the lightcurve. Performing a Lomb-Scargle (\citealt{lomb,scargle}) analysis on the lightcurve revealed a high power at a frequency around 550 seconds ($\approx$ 158 c/d). The high significance of the power was also tested via Monte-Carlo simulations, leading to a significance detection above 99.99\%. The observed power spectrum is shown in Fig. \ref{fig:2}. 

From the above described observations and analysis we can conclude that RXJ1654 displays a periodic signal compatible with that of a spinning magnetised WD. The result agrees with that of \cite{lutovinov}, but this is still not sufficent to firmly confirm the nature of RXJ1654 as an IP. Because of this we decided to begin a global observation campaign to obtain photometry over a longer timeline on the source to observe possible orbital modulations and possible orbital sideband modulations as well. This is described in the next section.  

%

\section{CBA photometry observations}
The optical counterpart of RXJ1654 was bright enough to be observable with the small-telescope network of the Center for Backyard Astrophysics (CBA: \citealt{skill_patt}), and the star's favourable declination made it accessible from all observing stations.  Thus we were able to obtain time-series differential photometry spanning a full range of terrestrial longitude, with coverage sometimes approaching 24 hours/day. Our 27-day campaign in April/May 2010 from nine telescopes included 37 nightly time series spanning 191 hours with a cadence ranging from 24 to 70 seconds.  A summary of the observations is presented in Table \ref{table:1}.

\begin{table}
\caption{Global time series photometry.}
\begin{tabular}{c c c}
\hline
 Telescope & Observer & Night/Hours \\[0.5ex]
\hline
CBA-Nelson 0.35 m    &  R. Rea      &     9/60 \\
CBA-Pretoria 0.35 m  &  B. Monard   &     5/36 \\
CBA-New Mexico 0.3 m &  T. Krajci   &     7/32 \\
MDM 1.3 m            &  H. Uthas    &     4/16 \\
CBA-Pukekohe 0.25 m  &  S. Lowther  &     3/13 \\
CBA-Perth 0.35 m     &  G. Bolt     &     2/10 \\
CBA-Nomad 0.4 m      &  G. Myers    &     5/6  \\
CBA-Alfred 0.8 m     &  R. Link     &     1/3  \\
CBA-Arkansas 0.4 m   &  T. Campbell &     1/4  \\

\hline 
\end{tabular}
\\ 
\label{table:1}
\end{table}

\begin{figure}
\includegraphics[width=0.5\textwidth,  height=0.3\textheight]{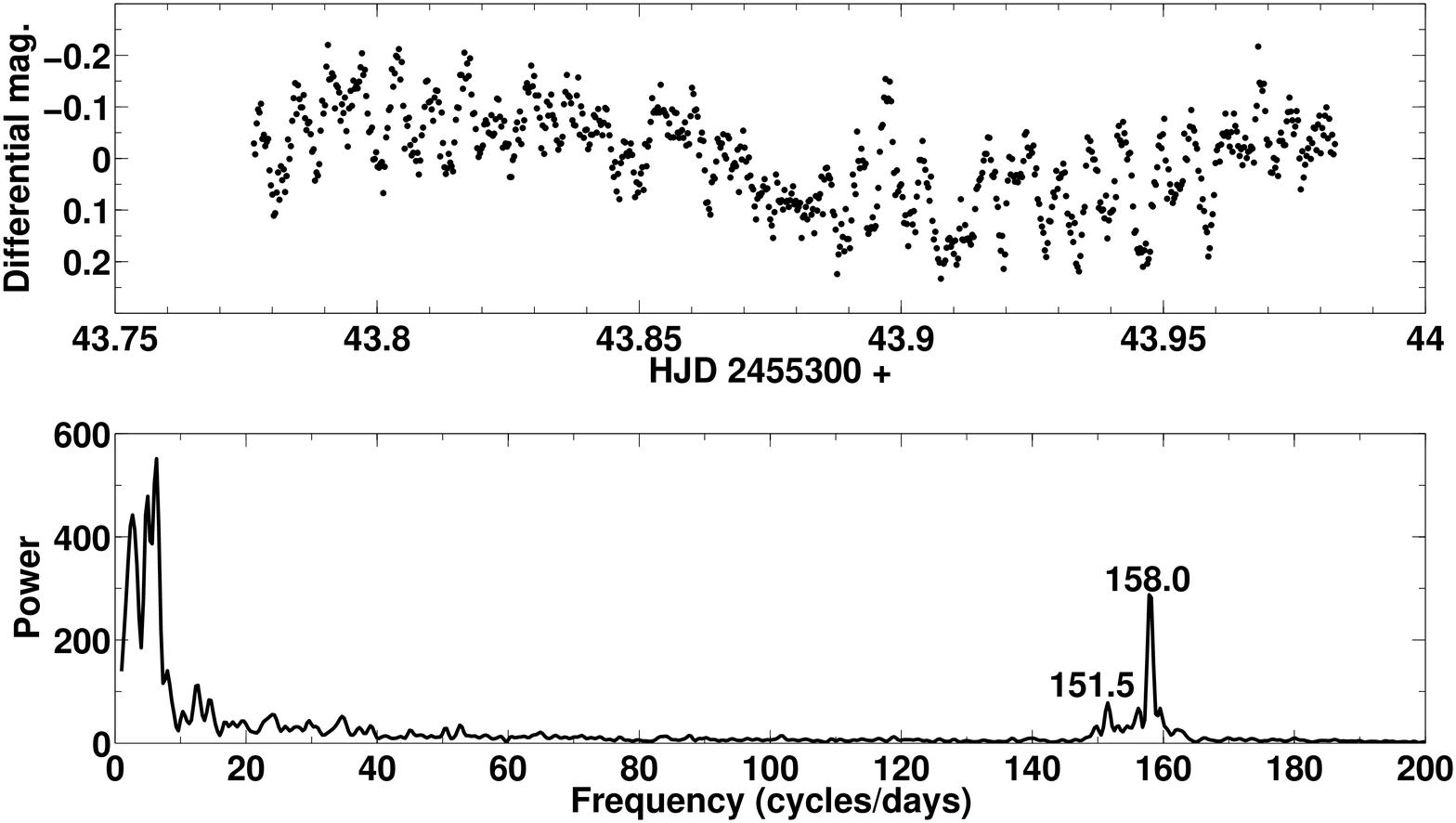}	
\caption{Upper frame: a sample nightly light curve, at 24 s/point. A nine minute pulse of varying amplitude is visible. In retrospect, the amplitude changes appear to be almost entirely caused by the beating of the signals closely spaced in frequency. Lower frame: average power spectrum of the eight best nights, revealing periodic signals, flagged with their frequency in cycles/day (with $\pm$0.2 errors).  The broad excess of power near 155 c/d is entirely owing to the varying power-spectrum window of the eight nightly time series.}
\label{fig:3}
\end{figure}

One night's lightcurve is shown in the upper frame of Fig. \ref{fig:3}. Direct inspection reveals its main features: rapid fluctuations at a period near nine minutes, and a slow wave on a roughly four-hour timescale. All lightcurves looked essentially like this, and the mean brightness was consistently near $V=15.6\pm0.2$. We selected the eight best nights (each comprising 6-19 hours of good data) and calculated their individual power spectra with a discrete Fourier transform.  They all look very similar, and the lower frame of Fig. \ref{fig:3} shows the average. The sharp peaks indicate periodic signals, labelled with their frequency in cycles/day. There is a poorly-defined excess of power at low frequency (above the general rise due to red noise); but this appears to be mostly owing to a weak signal near 6 c/d. No significant power is seen at any other frequency, up to 2000 c/d. The general appearance of the power spectrum and lightcurves is typical of intermediate polars (\citealt{patterson}).

\begin{figure}
\includegraphics[width=0.5\textwidth, height=0.25\textheight]{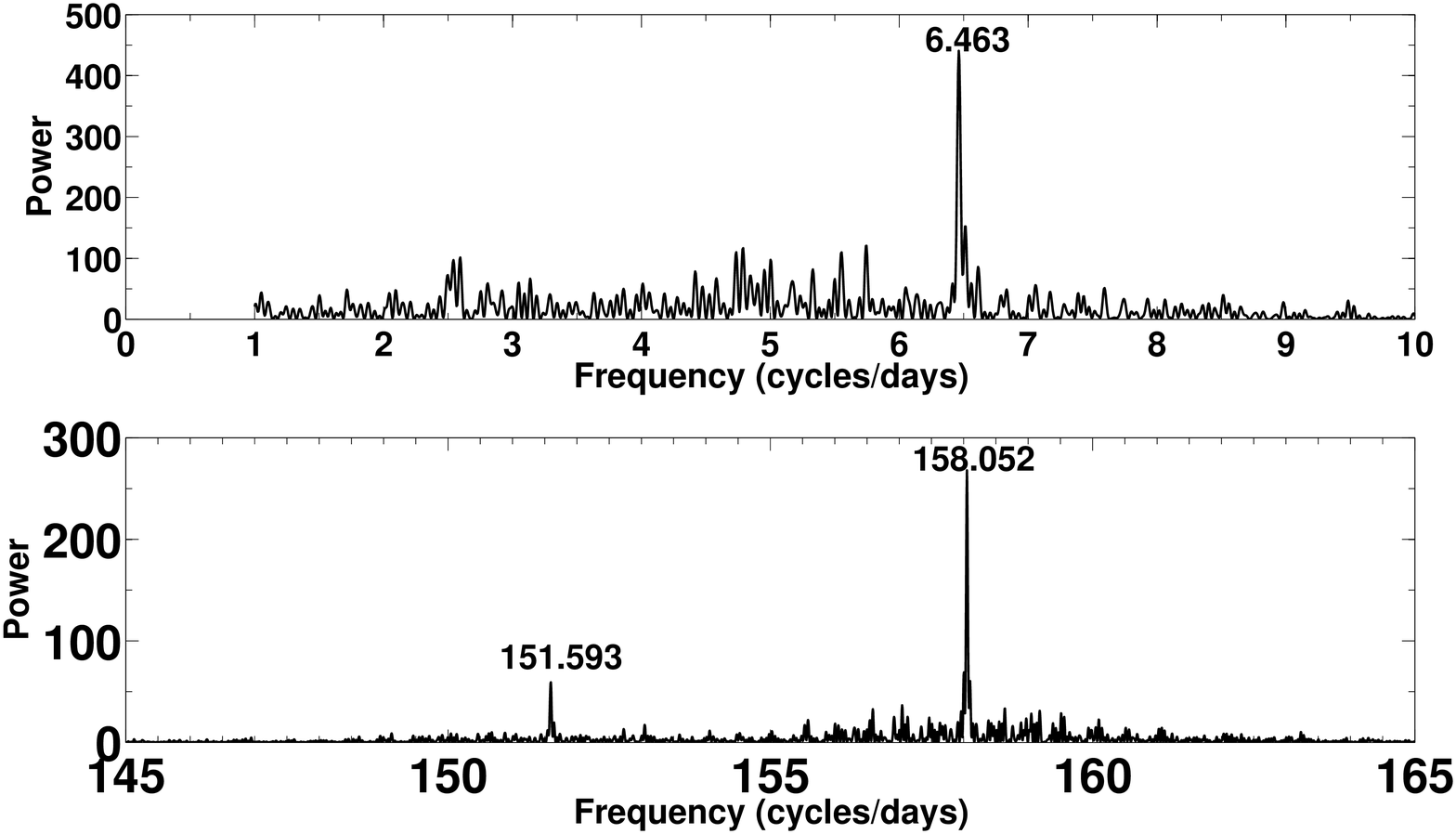}	
\caption{Segments of the high-resolution (27-night) power spectrum, showing that the light curve is dominated by three periodic signals, flagged with their frequency in cycles/day ($\pm$0.002). The low frequency is apparently the beat (``orbit'') of the two high frequencies (``spin'' and ``sideband'').}
\label{fig:4}
\end{figure}

Critical for that interpretation, however, is the coherence of the periodic signals. Because the star showed only small night-to-night changes in brightness, we subtracted the mean differential magnitude from each time series, thus ``zeroing'' each night. This removes the inevitable and undesirable small calibration offsets among the many telescopes, but does not result in the significant loss of information. We spliced all the zeroed time series of acceptable quality (these 8, plus 18 others which were briefer or of somewhat lower quality) and calculated the power spectrum of the entire time series, which spanned 27 nights (JD 2455340-2455367, or 340-367 in the shorthand we will adopt from here on). The resulting power spectrum is shown in Fig. \ref{fig:4}, and revealed narrow spikes at 6.463, 158.051, and 151.593 (each $\pm0.002$) c/d. The widths of the peaks were consistent with the signals' full stability - constant amplitude and phase throughout the 27 nights. This is ``standard fare'' for intermediate polars, where the signals are interpreted respectively as the orbital frequency $\Omega$, the spin frequency $\omega$, and the $\omega-\Omega$ sideband, which may arise from the reprocessing of the spin-modulated flux by structures fixed in the orbital reference frame.

\begin{figure}
\includegraphics[width=0.5\textwidth, height=0.25\textheight]{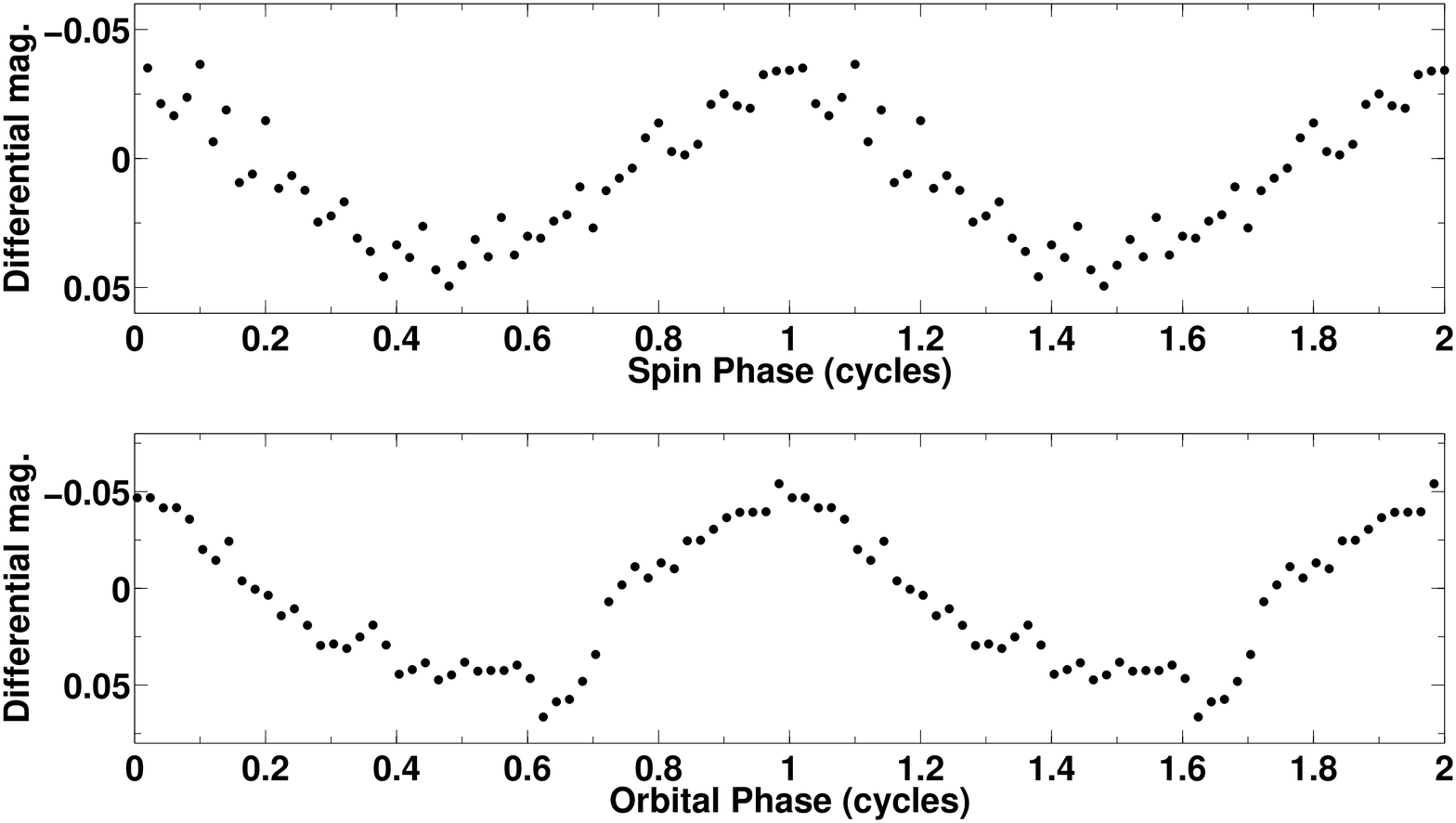}	
\caption{Mean waveforms of the spin and orbital signals.}
\label{fig:5}
\end{figure}

The global span of the observations (South Africa, North America, New Zealand, western Australia) suppressed aliasing, as evidenced by the lack of sideband peaks displaced by $\pm1$ c/d. The mean waveforms of orbital and spin signals are shown in Fig. \ref{fig:5}. The spin, orbital and $\omega-\Omega$ signal (not shown) signals are essentially a perfect sinusoid, with the ephemeris summarised in Table \ref{table:2}. A possible 0.03 mag ``dip'' near the orbital phase of 0.66 is visible. The two high-frequency signals evidently line up in phase at orbital phase 0.73(4).

\begin{table*}
\caption{Periodic signal properties, with uncertainties quoted in parenthesis in units of the last quoted digit.}
\begin{tabular}{c c c c}
\hline
Period & Semi-amplitude & Maximum light & Interpretation\\[0.5ex]
(s)    & (mag)          & (HJD+2455000) & \\[0.5ex]
\hline
546.6606(17) & 0.034(2) & 341.82943(7) & Spin ($\omega$) \\
569.955(13)  & 0.015(2) & 349.0353(3)  & Sideband ($\omega-\Omega$) \\
13375(4)     & 0.047(3) & 347.542(3)   & Orbit ($\Omega$) \\
\hline 
\end{tabular}
\\ 
\label{table:2}
\end{table*}

\begin{figure}
\includegraphics[width=0.5\textwidth]{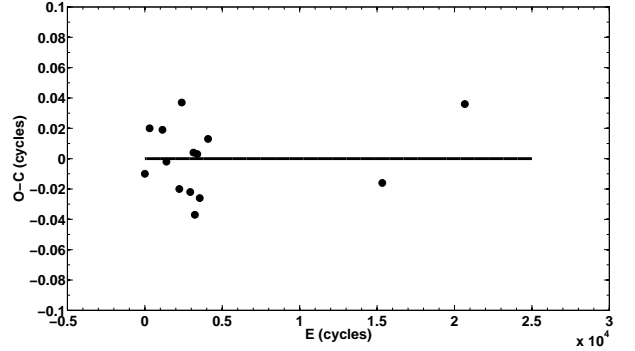}	
\caption{O-C diagram of the spin-pulse maxima given in Table 3 with respect to Eq. \ref{eqn:4}.}
\label{fig:6}
\end{figure}

The spin signal was sufficiently powerful to yield a significant pulse timing on each night. These timings are presented in Table \ref{table:3}; they span the main cluster of time series during JD 340-367, plus a few short time series much later in the observing season. A linear fit to these timings (with low weight accorded to the two brief late-season timings) yields an ephemeris 
\begin{multline}
\text{HJD}_{\rm max} =\\ \text{ HJD 341.82943(7) + 0.00632709(2)$\times$E}\label{eqn:4}
\end{multline}
The corresponding observed - computed (O-C) ephemeris from Eq. \ref{eqn:4} is displayed in Fig. \ref{fig:6}. The ephemeris spans 131 days, and should be sufficiently accurate to enable a future cycle count from year to year.

\begin{table}
\caption{From top to bottom, left to right: nightly maximum light of the white dwarf spin period of 546 seconds (HJD 2455000+).}
\begin{tabular}{| c | c | c |}
\hline
 341.8294  &   343.7782  &   349.0297 \\
 350.6556  &   355.9133  &   356.1479 \\
 356.9197  &   360.3739  &   361.6901 \\
 362.2656  &   363.2592  &   364.2587 \\
 367.6566  &   438.8615  &   472.5789 \\

\hline 
\end{tabular}
\\ 
\label{table:3}
\end{table}

\section{Time-resolved spectroscopy}
We obtained time series spectra of RXJ1654 in 2010 June, using the 2.4 m Hiltner Telescope at MDM Observatory, on Kitt Peak, Arizona (USA). All the spectra were taken with the Modular spectrograph. A 600 line mm$^{-1}$ grating and SiTE CCD detectors yielded 3.9 \AA\ resolution (FWHM) and 2.0 \AA / pixel$^{-1}$. The first twenty-one spectra were obtained  on June 15 and 16 UT, using a 2048x2048 detector that covered from 4210\AA\ to 7500\AA\ (with severe vignetting towards the ends). A fault in this detector forced a change to a similar 1024x1024 CCD for the last three spectra (obtained June 18, 19, and 20); these covered from 4660\AA\ to 6730\AA. Nearly all exposures were 10 min long; a few were 8 min. We pushed the observations to large hour angles to defeat daily cycle count ambiguities in the resulting velocity time series and rotated the instrument as needed to position the spectrograph slit near the parallactic angle (\citealt{filippenko82}).

We observed flux standards, and the weather was clear, but uncalibratable losses at the $\sim 1$ arcsec slit limit the  absolute calibration to $\sim 20$ \% accuracy.  The shape of the continuum is not well-constrained at the ends of the spectral coverage.  Comparison spectra obtained in twilight yielded fits with RMS pixel-to-wavelength residuals of $\le 0.05$\AA. The wavelength zero point for each programme object spectrum was derived from the 5577\AA\ night-sky feature. 

\begin{figure}
\includegraphics[width=0.45\textwidth]{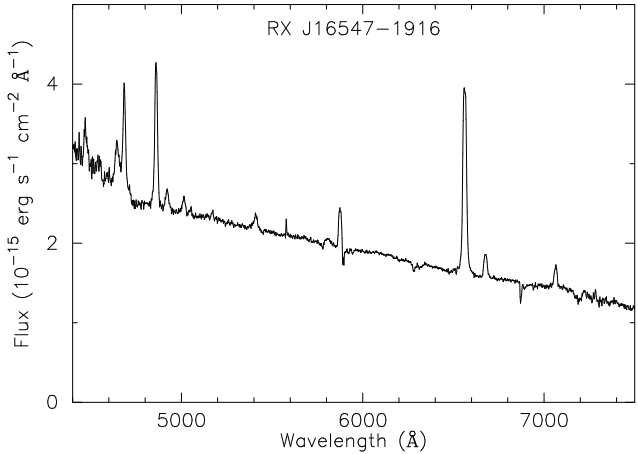} 	
\caption{Average spectrum observed in 2010 June. The vertical scale is uncertain by approximately 20 per cent.}
\label{fig:7}
\end{figure}

Fig. \ref{fig:7} shows the mean flux-calibrated spectrum. The emission lines (listed in Table \ref{table:4}) are for the most part Balmer, He I, and He II features typically found in DQ Her/IP stars. The CIII-NIII blend near 4640\AA\ is also detected. The Na D lines appear sharp, indicating interstellar origin, and the absorption near 6280\AA\ may include a contribution from the diffuse interstellar band (\citealt{jenniskensdesert}). The continuum follows an approximate power law, $f_\lambda = \lambda^{-1.5}$. The flux level corresponds to $V = 15.7$, using the $V$ passband tabulated by \cite{bessell90}. 

\begin{table}
 \centering
 \begin{minipage}{140mm}
  \caption{Mean spectrum emission lines.}
  \begin{tabular}{@{}lrrr@{}}
  \hline
  Line & E.W.\footnote{In Angstroms; emission counted as positive} & 
   Flux\footnote{In units of $10^{-16}$ erg s$^{-1}$ cm$^{-2}$.}  & 
   FWHM\footnote{In Angstroms; from Gaussian fits.} \\
\hline
CIII-NIII $\lambda 4647$ & $  7$ & $186$ & 33 \\ 
 HeII $\lambda 4686$ & $ 11$ & $286$ & 22 \\ 
            H$\beta$ & $ 15$ & $356$ & 17 \\ 
  HeI $\lambda 4921$ & $  2$ & $ 54$ & 19 \\ 
  HeI $\lambda 5015$ & $  2$ & $ 41$ & 21 \\ 
   Fe $\lambda 5169$ & $  1$ & $ 13$ & 14 \\ 
 HeII $\lambda 5411$ & $  3$ & $ 56$ & 28 \\ 
  HeI $\lambda 5876$ & $  5$ & $ 91$ & 17 \\ 
           H$\alpha$ & $ 39$ & $626$ & 22 \\ 
  HeI $\lambda 6678$ & $  4$ & $ 65$ & 20 \\ 
  HeI $\lambda 7067$ & $  4$ & $ 51$ & 18 \\ 
\hline
\end{tabular}
\end{minipage}
\label{table:4}
\end{table}

\begin{figure}
\includegraphics[width=0.45\textwidth]{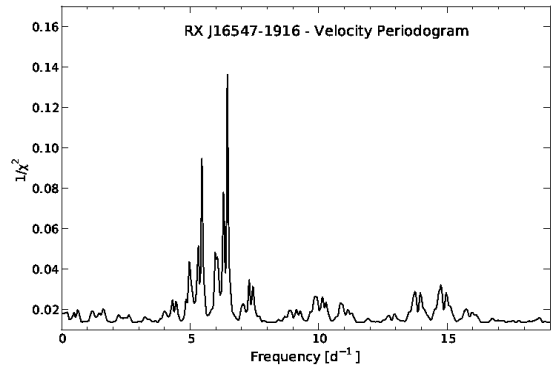}	
\caption{Period search of the H$\alpha$ radial velocities.}
\label{fig:8}
\end{figure}

\begin{figure}
\includegraphics[width=0.50\textwidth]{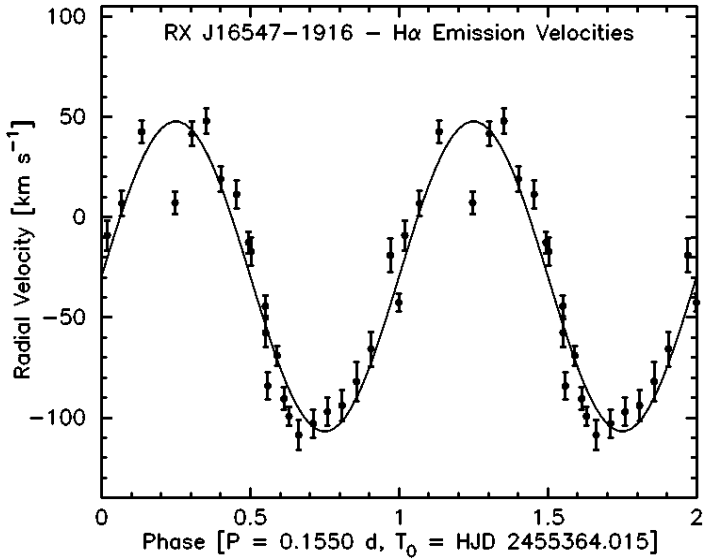}	
\caption{Radial velocities of the H$\alpha$ line, folded on the best-fitting period. All data are shown twice for continuity, and the sinusoidal fit is superposed.  The error bars are estimated  from counting statistics. The zero point of the phase in this figure corresponds to phase $0.42 \pm 0.04$ in the photometric orbital ephemeris. }
\label{fig:9}
\end{figure}

We measured velocities of the strongest line (H$\alpha$) using a convolution algorithm described by \cite{sy80} and \cite{shaf83}. The convolution function consisted of a positive and negative Gaussian of FWHM 12 \AA, with centre-to-centre separation of 28 \AA; these values were chosen to emphasise the steep sides of the line profile and obtain adequate signal-to-noise. This yielded the velocity time series given in Table \ref{table:6}.  We searched for periods in the time series using the least-residuals algorithm described by \cite{tpst}, and found a strong periodicity near 6.45 c/d (see Fig. \ref{fig:8}). Fig. \ref{fig:9} displays the H$\alpha$ radial velocities folded on the best-fitting period. A fit near this frequency, of the form  $v(t) = \gamma + K \sin [2 \pi (t - T_0) / P]$ returns the values given in Table \ref{table:6}, where the uncertainties (in parentheses, in units of the last quoted digits) are derived from the scatter in the best fit; $\gamma$ especially, and $K$ as well, are most probably subject to systematic uncertainties larger than their statistical errors. We note that the radial-velocity period, 0.1550(2) d, is identical (within its uncertainty) to
the more precise photometric orbital period found in Section~3.

The velocity near spectroscopic phase 0.22 is lower than expected (Fig. \ref{fig:9}). Spectroscopic phase 0.22 corresponds to a phase of $0.42 \pm 0.04$ in the photometric ephemeris, so this point lies close to the phase of the 0.03 mag dip referred to in Section~3. The available data have no other velocities at this phase, so we cannot establish whether this effect is reproducible.

\begin{table*}
 \centering
 \begin{minipage}{140mm}
  \caption{H$\alpha$ radial velocities.}
  \begin{tabular}{| crr | crr | crr |}
  \hline
  $t$\footnote{HJD minus 2455360.} & $v(t)$ & 
   $\sigma$\footnote{Estimated from counting statistics.} &
  $t$ & $v(t)$ & $\sigma$ &
  $t$ & $v(t)$ & $\sigma$ \\
    & (km s$^{-1}$) & (km s$^{-1}$) & 
    & (km s$^{-1}$) & (km s$^{-1}$) & 
    & (km s$^{-1}$) & (km s$^{-1}$) \\ 
\hline
   2.8602 & $  -44$ &    5  &    3.7006 & $  -19$ &    8  &    3.9148 & $   48$ &    6  \\
   2.8664 & $  -69$ &    5  &    3.7081 & $   -9$ &    7  &    3.9223 & $   19$ &    6  \\
   2.8726 & $  -99$ &    5  &    3.7157 & $    7$ &    6  &    3.9305 & $   11$ &    7  \\
   3.6603 & $ -103$ &    7  &    3.7260 & $   43$ &    5  &    3.9381 & $  -17$ &    7  \\
   3.6678 & $  -97$ &    7  &    3.7817 & $  -13$ &    5  &    3.9456 & $  -58$ &    7  \\
   3.6754 & $  -94$ &    8  &    3.8002 & $  -91$ &    6  &    5.7589 & $    7$ &    6  \\
   3.6829 & $  -82$ &   10  &    3.8078 & $ -109$ &    8  &    6.8057 & $  -43$ &    4  \\
   3.6905 & $  -66$ &    9  &    3.9072 & $   42$ &    6  &    7.8227 & $  -84$ &    7  \\
\hline
\end{tabular}
\end{minipage}
\label{table:5}
\end{table*}

\begin{table}
\caption{Best-fit parameters of $v(t) = \gamma + K \sin [2 \pi (t - T_0) / P]$ to Fig. \ref{fig:8}, with uncertainties quoted in parenthesis in units of the last quoted digit.}
\begin{tabular}{l l}
\hline
T$_0$ & HJD 55364.0152(19) \\
P & 0.1550(2) \\
K & 77(8) km s$^{-1}$ \\
$\gamma$ & -30(5) km s$^{-1}$\\ 
\hline 
\end{tabular}
\\ 
\label{table:6}
\end{table}

\section{Discussion}

The results presented here confirm the true nature of RXJ1654 as an IP. When viewed in the optical light the observed features of this source are quite typical of the IP class (\citealt{warner,patterson}), displaying photometric modulations on the orbital and spin period, radial velocity shifts on the orbital timescale and a very ``standard'' mean spectrum. However, when viewed through the X-ray light, this system is part of a smaller subset of hard X-ray emitting IPs. As described in \cite{scaringi}, IPs identified through optical follow-up observations of unidentified hard X-ray sources all have spin periods below 30 minutes and spin-to-orbital ratios below 0.1 \footnote{with the only exception of the close by IP EX Hya}, and RXJ1654 is no exception to this with a $\omega/\Omega$ ratio of about 0.04. 

We might try to understand the relative phasing of the two orbital clocks (the radial velocities and the photometric wave) by making two very optimistic assumptions: the radial velocities signify motion of the white dwarf, and the photometric wave signifies heating of the secondary. On these assumptions, we expect maximum light at the time of blue-to-red crossing.  But on the ephemeris of Table \ref{table:2}, maximum light occurs at spectroscopic phase 0.58 (i.e. $0.58 \pm 0.04$ cycles too late). We conclude that one or both of these optimistic assumptions must be false. It is worth noting however that the Balmer lines are known to display different contributions (\citealt{hellier,zhar}). It could be that the radial velocity curve, which refers to the wings of the H-$\alpha$ line, originates in the accretion regions closest to the WD. Thus, unless the line wings are contaminated by other contributors, the phase shift between optical light and radial velocity could suggest that the optical light is not dominated by the secondary, but instead by a region in the outer edge of the accretion disc, such as a hot-spot.
 
Given that RXJ1654 is a fairly typical IP of the ones observed in the hard X-ray regime, it is possible that the WD is in spin equilibrium, displaying a mixture of disk/stream accretion, whilst also propelling material away. If this were the case, we would expect to observe X-ray modulations on the spin period and moreover we would also expect the spin frequency of the WD to deviate from the current value with a spin-up/down timescale $\leq10^{7}$ years (\citealt{norton_theo1,norton_theo2}). However, not all hard X-ray emitting IPs need to be in equilibrium, and in fact the hard X-ray emitting AE Aqr is in a purely propellor stage, and has been showing rapid spin-down rates (\citealt{wynnAE,dejager}), as the WD spin tries to reach equilibrium for its given mass ratio and magnetic field strength. It would be interesting to follow-up on RXJ1654 and other hard X-ray IPs to determine whether any indication of spin-up/down of the WD is observed. If we take the \cite{norton_theo1} models, the combination of spin and orbital period together with the mass ratio and spin up/down rates could yield an indication of the WD magnetic field strength. However, all the information required to provide such an estimate is not available to date, and particularly the mass ratio of this system is not know, which leads to an estimate of the magnetic field strength spanning over an order of magnitude (Fig. 4 from \citealt{norton_theo1}).

To date, more than 30 mCVs have been observed above 15 keV. Of these, about eight are new discoveries from optical follow-up observations of unidentified sources in the {\it INTEGRAL}/IBIS survey catalogues (\citealt{cat4} and previous catalogues). Moreover, there is an additional similar number of systems identified as being CVs through optical spectroscopy. That no one has yet obtained optical photometry inhibits the determination of whether these systems are IPs or not. As shown by \cite{pretorius}, care needs to be taken when trying to infer whether a candidate hard X-ray emitting CV contains a magnetic WD, or not. In fact, in one case one of the systems was later found not to be a CV at all. On the other hand, however, the majority of hard X-ray candidate CVs do turn out to be IPs, with fast spinning WD, and spin-to-orbital ratios $\leq$0.1. We are therefore confident that many additional such IPs will be discovered as optical follow-up of unidentified hard X-ray sources becomes more complete and hard X-ray surveys push to increasing depth. In the latest {\it INTEGRAL}/IBIS catalogue, $\approx$9\% of the identified population has been confirmed as mCVs, while 30\% of the catalogued sources still remain to be identified from the whole survey catalogue ($\approx$20 new CVs should be discovered from the unidentified source population). The work of \cite{masettiVIII} (and previous related work) has spectroscopically identified about eight candidate CVs which still need photometric observations to confirm their magnetic nature. In particular all \textit{INTEGRAL} detected CVs comprise the small, but fast growing, ``\textit{INTEGRAL}-selected'' population of IPs. This is a cleanly selected sample of objects when compared to the general IP population, and we highlight that only clean samples like this can really be used to study the overall populations and evolution of magnetic CVs, and magnetic binaries in general.

\section{Conclusions} 

Here we have presented optical follow-up observations of the candidate CV source RXJ1654. We updated the nature of this system as another hard X-ray emitting IP, confirming the \cite{lutovinov} detection of a $\approx$550 seconds in the system as the WD spin signal. This finding adds to the small, but fast growing, subset of IPs all displaying persistent hard X-ray flux above 15 keV, spin periods below 30 minutes, and spin-to-orbital ratios below 0.1.

\begin{acknowledgements}
This research has made use of NASA's Astrophysics Data System Bibliographic Services. Thanks go to the nightly IAC80 support astronomers Joge Garcia, Christina Zurita and Santiago Lopez for the data acquisition. S.S. acknowledges funding from NWO project 600.065.140.08N306 to P.J. Groot. JP is gratefull for support from the NSF grant AST-0908363 and the Mount Cuba Astronomical Foundation. JRT gratefully acknowledges support from NSF grant AST-0708810. Special thanks also go to all the students and supervisors Tony Bird  and Ismael Perez Fournon, present at the 2010 Tenerife field trip organised by the University of Southampton and the University of La Laguna.
\end{acknowledgements}

\bibliographystyle{aa} 
\bibliography{RXJ1654} 

\end{document}